\documentclass[12pt,draftclsnofoot,onecolumn]{IEEEtran}
\usepackage{graphicx}
\usepackage{array}
\usepackage{amssymb}
\usepackage{setspace}
\usepackage{amsmath}
\usepackage{color}
\usepackage{subfig}
\usepackage{comment}
\usepackage{enumerate}
\usepackage{url}
\usepackage{multirow}
\usepackage{footnote}
\usepackage{breqn}
\usepackage{lmodern}
\usepackage[T1]{fontenc}

\begin{document}

\doublespacing

\title{Downlink Throughput Driven Channel Access Framework for Cognitive LTE Femto-Cells}

\author{Mohamed Hamid, Slimane Ben Slimane, and Niclas Bj\"orsell}
\date{}

\maketitle
\begin{abstract}

This paper proposes an optimized sensing based channel access framework for the LTE cognitive femto-cells, with an objective of maximizing the femto-cells downlink throughput. Cognitive femto-cells opportunistically transmit on the macro-cell channels when they are free of use. Those free channels are located by means of spectrum sensing using energy detection. Moreover, periodic sensing is adopted to detect any changes of the sensing outcomes. The maximum attainable femto-cell downlink throughput varies with the macro-cell channel occupancy statistics. Therefore, the LTE macro-cell occupancy is empirically modeled using exponential distributions mixture. The LTE cognitive femto-cell downlink throughput is maximized by compromising the transmission efficiency, the explored spectrum opportunities and the interference from the macro-cell. An analytical solution for the optimal periodic sensing interval that maximizes the throughput is found and verified by simulations. The obtained results show that there is indeed a single periodic sensing interval value that maximizes the LTE cognitive femto-cell downlink throughput. At the peak of the macro-cell traffic, our framework increases the femto-cell throughput by $\simeq 15\%$ compared to the senseless case. The impact of the available number of channels for opportunistic access is studied and no significant impact is found for more than three channels.
\end{abstract}


\section{Introduction} \label{sec:intro}


Mobile operators are facing an explosive growth in data traffic which imposes a great challenge on how to be handled with the available radio resources. Many studies have shown that this huge data traffic is mostly indoor originated \cite{FBS_survey}. Therefore, a promising solution for this challenge is to approach towards more distributed networks architecture using low power short range indoor access points. In this regard, the third generation partnership project (3GPP) provides standards for next generation cellular systems that support deployment of reduced scale plug and play access points. These access points are known as femto-cell base stations which are connected to the mobile core network through internet cloud. \cite{femto_stand}. Femto-cell base stations together with the macro-cell base station (MBS) serving the same geographical area form a network topology termed as two tier heterogeneous network.

From radio network planning prospective, femto-cell base stations can share the spectrum owned by the MBS on an opportunistic base under cognitive radio (CR) framework \cite{Mitola}. With CR, secondary unlicensed users can access the radio channels when they are not being utilized by their primary licensed users. These free radio channels are called spectrum opportunities \cite{Haykin}. Therefore, letting the femto-cell base stations to take the role of secondary users and the MBSs to act as primary users, would facilitate achieving higher network throughput with more efficient use of the licensed spectrum owned by the network operator \cite{cognitive_LTE_femto}. Femto-cell base stations sharing the spectrum with MBSs are called cognitive femto-cells. The potentials of cognitive femto-cells are extensively investigated in the literature, for example, see \cite{femto_lit_1, femto_lit_2, femto_lit_3}. As a 3GPP standard, LTE femto-cell base stations deployment is supported which will be referred to as FBS hereafter. In this paper, an optimized channel access framework for the FBSs that maximizes their throughput is proposed. Below are the connections between what has been done in the literature and the contributions of this paper.

To the best of the authors' knowledge, it has always been assumed that an accurate spectrum sensing is performed by the FBS \cite{cognitive_LTE_femto,interf_mgnt}. Based on that assumption, FBSs are considered to continuously have perfect knowledge regarding the spectrum opportunities availability. This assumption is reasonable if the concern is the instantaneous status of a specific channel at sensing especially with considering the high transmission power of the MBS that eases their detectability. However, there is no guarantee that this status will hold to the end of using the channel by the FBS or to the next sensing as the MBS may resume its transmission. Consequently, MBS activity statistics directly influence the interfered portion of FBS transmission. Hence, obtaining these statistics of MBS activity is an important prerequisite for system's parameters setting. In this regard, many studies have been carried out to characterize the downlink cellular channel occupancy statistical distribution. In \cite{Rap}, it is shown that exponential distribution can approximate mobile telephony channel occupancy with its advantage of traceability in finding analytical solutions for optimization problems. Therefore, exponential distribution has been intensively used to model cellular channel occupancy, see \cite{optimal_sensing, Hamidcogart, Hamidcmc} as examples. Nevertheless, many empirical studies have shown a heavy tail behaviour for the cellular channels occupancy which is not properly characterized by exponential distributions \cite{chn_ocup_exp_ex}. Accordingly, some other heavy tail distributions are used as alternatives to model the cellular channel occupancy, among which are log-normal, \cite{Aachen_dist,chn_ocup_lgn_ex}, Beta and Kumaraswamy distributions \cite{Lopez_dist}. Aiming at preserving the advantage of exponential distributions traceability and characterizing the heavy tail behaviour of cellular channel occupancy at the same time, in this paper, exponential distributions mixture is used for LTE MBS downlink channel occupancy modeling. Using of exponential distributions mixture is firstly used for tele-traffic modeling in \cite{hyper_exp} and adopted in \cite{Luca_ex_dist} to model channel occupancy in WiFi systems. Moreover, using exponential distributions mixture is motivated by the fact that combining multiple distributions linearly improves the goodness of fitting compared to the case of single distribution fitting. 

Primary system channel occupancy model based optimization of sensing parameters has been an attractive research problem in CR arena. In \cite{optimal_sensing} the objective of the sensing parameters optimization is set as to maximize the transmission efficiency subjected to interference mitigation. In \cite{Hamidcogart}, mutual iterative optimization algorithm for sensing time and periodic sensing interval is introduced. The algorithm presented in \cite{Hamidcogart} optimizes the sensing parameters from sensing reliability prospective and no interference or throughput considerations are taken into account. In \cite{Hamidcmc}, a multi-channel system is considered for periodic sensing intervals optimization. The driving objectives in \cite{Hamidcmc} are to maximize the utilization of the available spectrum opportunities and to minimize the idle channel search delay. Sensing time is optimized in \cite{sensing_thrput} with an objective of secondary system throughput maximization constrained by primary system protection. In \cite {sensing_thrput}, no channel occupancy model is considered. Instead, the instantaneous throughput when a free channel is available is optimized. In contrast to \cite{optimal_sensing, Hamidcogart, Hamidcmc, sensing_thrput}, this paper tackles the sensing parameters optimization from different prospectives and addresses some of the simplifications and general assumptions carried out in the previous related work. At first, a specific scenario of dynamic spectrum sharing is considered, that is, LTE FBSs accessing the spectrum used by the MBSs opportunistically. By assuming this specific scenario, the general assumptions which affects the optimization procedure are eliminated. Moreover, more accurate results are obtained by using the actual transmission and deployment parameters for the specific sharing system. For example, in \cite{optimal_sensing, Hamidcogart, Hamidcmc}, the starting point is an assumption of a specific value of the received signal-to-noise-ratio (SNR). In this paper however, the deployment scenario, transmission power and the propagation models for LTE systems specified in the 3GPP standards are used. Furthermore, a simplified model of exponential distribution for the licensed user channel occupancy is used in \cite{optimal_sensing, Hamidcogart, Hamidcmc} while, in this paper, the statistical model for the channel occupancy is empirically developed, i.e., mixture of exponentials distributions. Furthermore, another added value for this paper is a study of the changes of the throughput with a real-life traffic variations.

To summarize, the distinct contributions of this paper are
\begin{enumerate}
\item Building an empirical statistical model for a primary system to facilitate the secondary system channel access optimization.
\item Considering a specific opportunistic spectrum access scenario of FBSs sharing spectrum with MBSs.
\item Proposing and evaluating an FBS throughput maximization based spectrum sensing framework.
\end{enumerate}

The rest of this paper is organized as follows: Section \ref{sec:model} presents the system model including deployment model, opportunistic spectrum access model and energy detection. The empirical model for the downlink LTE channel occupancy is presented in Section \ref{sec:modeling}. In Section \ref{sec:thrput}, the throughput based sensing optimization framework is explained. The obtained numerical results with their interpretations are included in Section \ref{sec:res}. Finally, Section \ref{sec:conc} provides the concluding remarks of the paper.

\section{System Model}\label{sec:model}
In this section the opportunistic spectrum sharing model adopted by the FBS is introduced. Besides, the two state Markov channel occupancy model is explained. Moreover, energy detection (ED) preliminaries are presented in this section.

\subsection{Deployment and sharing model}
\begin{figure}[t!]
  \centering
  \includegraphics[width=0.49\textwidth]{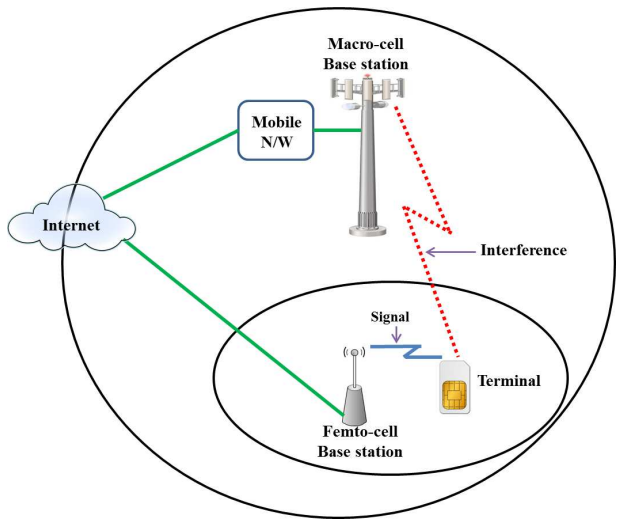} 
  \caption{Deployment and spectrum sharing model.}
  \label{fig:sys_model}
  \end{figure}

The FBS opportunistically accesses the same downlink spectrum assigned for the MBS which is divided into $L$ channels. The FBS is responsible for detecting the free channels available for its usage with no assist from the MBS. Moreover FBS checks for MBS transmission reappearance using periodic sensing. In case of MBS active transmission found, FBS releases the channel and starts looking for another free channel. Otherwise, FBS operation on the same channel continues. Due to this discrete sensing, there are occasions of FBS and MBS simultaneous transmission. This concurrent transmission is assumed to result in an interference from the MBS into FBS but not in the other direction under the assumption that MBS have considerably higher transmission power than FBS. Moreover, the lengths and amount of the instances of mutual transmission on a same channel are determined by the instances between the sensing occasions known as periodic sensing interval. However, assuming that the FBS can perform one task at a time, either sensing or communicating, then periodic sensing is an overhead that reduces the time during which the FBS can utilize free channels. Accordingly, finding the optimal periodic sensing interval is an optimization problem which can be objectively solved. The objective followed in this paper to handle this optimization problem is the FBS throughput's maximization. FBS throughput is chosen as an objective as the ultimate goal of FBS deployment is to provide high data rate indoor cellular link. Another assumption considered in this paper is the inter-FBSs interference which is assumed to be small and can be considered as a part of the background noise. Furthermore, due to their plug and play characteristics, it is reasonable to assume that FBSs have high duty cycle when they are active compared to the MBS. Therefore, for simplification and due to absence of FBS traffic models, one can assume that an FBS is actively transmitting as long as it is powered on. Fig. \ref{fig:sys_model} depicts the spectrum sharing model followed in this paper.

\subsection{Opportunistic spectrum access model}\label{sec:Markov}

Hereafter, the term channel will be used for an LTE MBS channel. From occupancy point of view, the channel can be modeled as a two state Markov process. These two states are: ON state representing occupied channel state and OFF state when the channel is idle. Each of ON and OFF states temporal length is a random variables (RV) with a specific statistical distribution. ON and OFF temporal length are assigned the RVs $x$ and $y$ respectively throughout the rest of this paper. The statistical distribution of $x$ and $y$ will be discussed in details with an empirical modeling in Section \ref{sec:modeling}. Channel utilization factor or duty cycle, $ u $, is defined as the fraction of time during which the channel is being utilized by its primary user, i.e., MBS in this paper. $u$ is mathematically obtainable as
  \begin{equation}
  u = \dfrac{E\{x\}}{E\{x\}+E\{y\}},
  \end{equation}  
  with $ E\{\cdot\} $ denoting the expected value.
\begin{figure}[t!]
    \centering
    \includegraphics[width=0.6\textwidth]{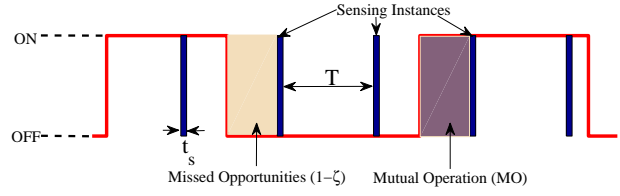} 
    \caption{Opportunistic channel access model.}
    \label{fig:ON_OFF_ilus}
    \end{figure}
The FBS senses a channel and starts to use it in case of MBS transmission absence. Otherwise, the FBS senses another channel. The sensing is performed periodically with a period of $T$ either to look for a free channel if none was found in the previous sensing or to detect MBS operation reappearance on the channel being used by the FBS. Transmission efficiency, $\eta$ is defined as the ratio between the time spent is utilizing a free channel, $T$, and the time spent on both utilizing the channel and sensing it. Therefore, $\eta$ is found as
  \begin{equation}
   \eta = \dfrac{T}{T + t_{s}}.
   \label{eq:eff}
   \end{equation}
where $t_{s}$ is the time required to perform the sensing called sensing time.

As the sensing is performed periodically in a discrete points in time, the following two situations are experienced
\begin{enumerate}
  \item The state of channel $l$ changes from ON to OFF state one or more times within a period of $ T $, meanwhile, the FBS captures a fraction of opportunities, call it $ \zeta $ and misses $ \left(1-\zeta\right)$ of the opportunities on that channel. A formula for finding $\zeta$ is provided in Subsection \ref{sec:hyper_exp} and Appendix A. For a system with $L$ channels, assume that the captured opportunities on channel $l$ is denoted as $\zeta_{l}$. Therefore, the whole system captured opportunities, $\zeta_{s}$, is obtained as
  \begin{equation}
  \zeta_{s} = 1-\prod\limits_{l=1}^{L}\left( 1 - \zeta_{l}\right).
  \label{eq:zeta_sys}
  \end{equation}

  \item  The state of channel $l$ changes from OFF to ON state one or more times within a period of $T$ while the FBS is utilizing it. Therefore, during a fraction of $T$ both the MBS and the FBS use the same channel. This fraction of time of mutual operation is derived in Subsection \ref{sec:thrput_infer}.
  
  \end{enumerate}
  
  The opportunistic channel access model for FBS with the captured/ missed opportunities and the mutual operation is depicted in Fig \ref{fig:ON_OFF_ilus} where the ON states are represented by the higher level of binary representation and the OFF states are represented by the lower state.
  
\subsection{Energy detection sensing} \label{sec:ED}
FBS performs spectrum sensing to locate a free channel. Spectrum sensing is basically the process of testing the existence of a primary user signal within a specific band or channel. For spectrum sensing, there are many proposed techniques in the literature which are surveyed in \cite{survey_2,survey_3,survey_1}. Among the proposed spectrum sensing techniques, ED is the simplest where received signal energy is compared with the background noise and accordingly primary user signal existence or absence is declared. The price of the ED simplicity is a degraded performance in low SNRs. However, in cellular systems and particularly in urban areas, the downlink SNRs are strong due to the high transmission power of the base stations. Hence, in accordance with this assumption, ED can be used as a reliable technique by FBS to detect the free channels.

When a channel is sensed, one of two hypotheses is declared: either the channel is free which is called the null hypothesis, $\mathcal{H}_{0}$, or the channel is used which is referred to as the positive hypothesis, $\mathcal{H}_{1}$. For the null hypothesis, noise only exists in the channel while for the positive hypothesis, noise bearing signal is received. Mathematically, the two hypothesis framework is expressed as
\begin{equation}
  r(t) = \left\{ 
    \begin{array}{l l}
    n(t) & \quad \mathcal{H}_{0} \\
     s(t)+n(t) & \quad \mathcal{H}_{1} \\ \end{array}, \right.
     \label{eq:2hypo}
  \end{equation}
where $r(t)$, $ n(t)$ and $s(t)$ are the received time domain signal, noise only components and signal components respectively. Accordingly, ED process is expressed as
\begin{equation}
   \left\{ 
    \begin{array}{l l}
 \sum\limits_{m=1}^{M} |r(m)|^{2} < \rho & \quad \mathcal{H}_{0} \\
     \mbox{Otherwise} & \quad \mathcal{H}_{1} \\ \end{array}, \right.
     \label{eq:ED}
  \end{equation}
where $m$ is an integer index for an $M$ discrete samples for the received signal. $ \rho $ is called the detection threshold which is obtained as explained later in this section. Assume that implementing (\ref{eq:ED}) is dominated by collecting the samples. Then by using Nyquist sampling criterion, the $M$ samples are collected in a time of $t_s$ where $M = 2 B t_{s}$, with $B$ denoting the channel bandwidth.
 
Two statistical performance metrics to evaluate spectrum sensing techniques are used, namely: the probability of false alarm, $ p_{fa}$, and the probability of detection, $p_{d}$. The probability of false alarm is the probability of wrongfully declaring signal existence while noise only is present. The probability of detection is the probability of detecting a signal that correctly exists. The probabilities of false alarm and detection are derived in \cite{optimal_sensing, Hamidcogart} as
  \begin{equation}
      p_{fa}= (1-u) \cdot Q  \left(\frac{\rho-2t_{s}B\sigma_{n}^{2}}{2\sqrt{t_{s}B}\sigma_{n}^{2}} \right),
      \label{eq:pfa}
  \end{equation}  
  \begin{equation}
      p_{d}= u \cdot Q \left(\frac{\rho-2t_{s}B(\gamma_{0}+1)\sigma_{n}^{2}}{2\sqrt{t_{s}B}(\gamma_{0}+1)\sigma_{n}^{2}} \right),
      \label{eq:pd}
  \end{equation}  
  where  $Q(\cdot)$ is the Q function representing the complementary cumulative distribution function (CCDF) of a Gaussian random process,  $\sigma_{n}^{2}$ is the noise variance and $\gamma_{0}$ is the received SNR. If the energy detector is set to achieve a specific probability of false alarm, then the detection threshold, $\rho$ is calculated as  
  \begin{equation}
       \rho = 2\sqrt{t_{s}B}\sigma_{n}^{2}Q^{-1}\left(\dfrac{p_{fa}}{1-u}\right)+ 2t_{s}B\sigma_{n}^{2},
      \label{eq:detection_threshold}
  \end{equation}
 where $Q^{-1}(\cdot)$ is the inverse Q function.

\section{Empirical LTE downlink channel Occupancy modeling}\label{sec:modeling}
This section presents the MBS channel occupancy statistical model developed as a part of this paper contribution. The section starts with introducing the measurements methodology and setup followed by the procedure of exponential distributions mixture fitting. At the end of the section, the fitting results are shown.

\subsection{Measurements}\label{sec:meas}

\begin{figure*}[h!]
  \centering
  \subfloat[]{\label{fig:meas_location}\includegraphics[width=0.4\textwidth]{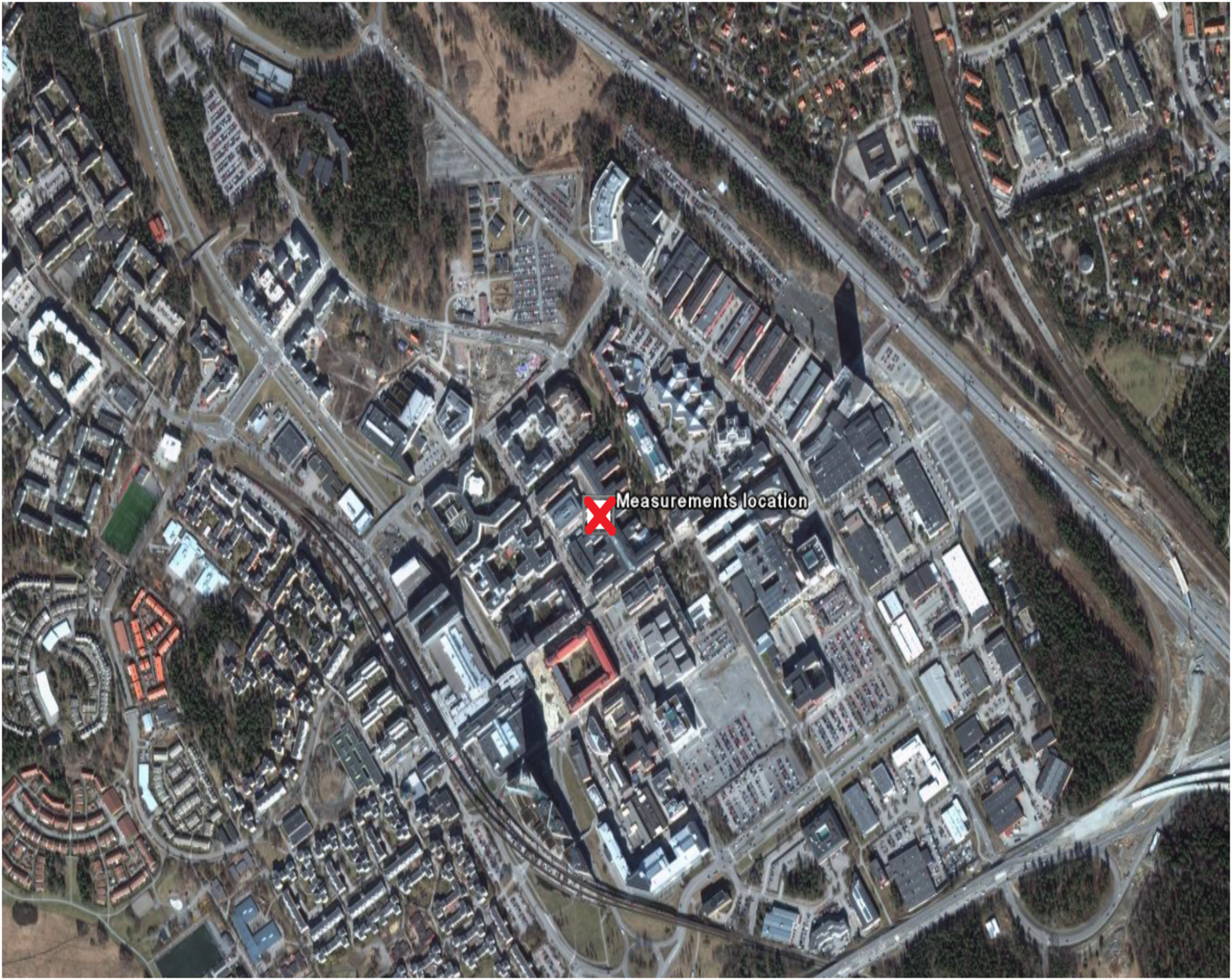}}
  \subfloat[]{\label{fig:meas_setup}\includegraphics[width=0.4\textwidth]{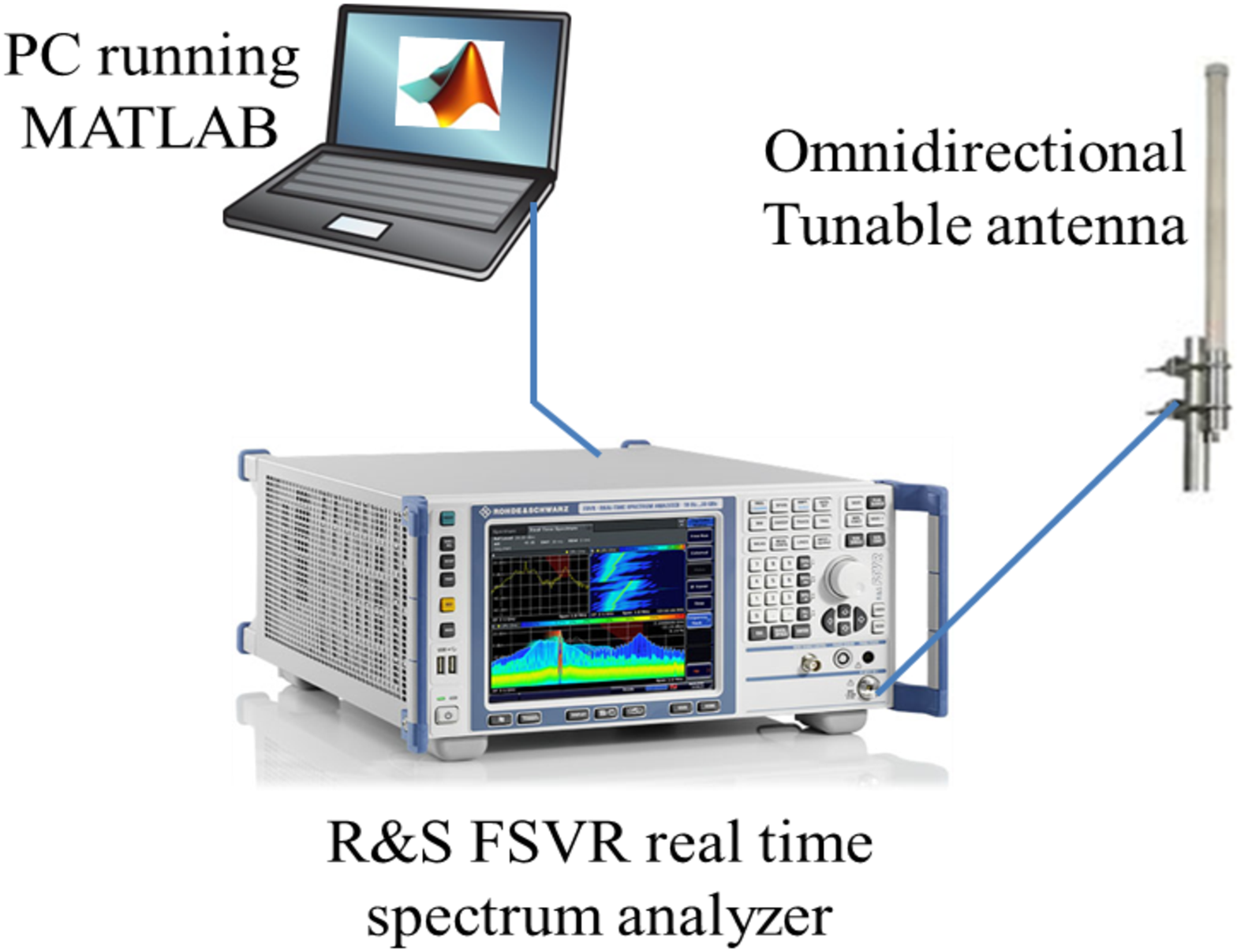}}
  
  \caption{  (a) Measurements location and (b) Measurements setup.}
  \label{fig:meas}
  \end{figure*}

To model the LTE downlink channel occupancy, a measurement campaign was held to fetch the empirical data. The measurements were held in an indoor location in Kista, Stockholm, Sweden in a university campus located at the GPS coordinates: $ 59^{0} \mbox{ }24^{'} \mbox{ }19.13^{"}\mbox{ N} $, $ 17^{0}\mbox{ } 56^{"}\mbox{ } 56.12^{"}\mbox{ E}  $.  The measurements location is located in one of the industrial centres of Stockholm where the surroundings are densely occupied by offices, a shopping mall and residential buildings. Fig. \ref{fig:meas_location} shows a Google map of the measurement location. The MBS signal is captured by an omnidirectional antenna connected to a real time spectrum analyzer (RTSA) to obtain a real time measurement data. The RTSA is controlled through a computer which is also used to record the data for further analysis. Fig. \ref{fig:meas_setup} shows the measurements setup. The measured MBS signal is $ 40 $ MHz width (the full RTSA bandwidth) centred at $ 2650.0 $ MHz. To cope with the variation of the traffic the collected measurements are divided into periods of four hours each. The RTSA measures the signal in the whole bandwidth every $30.0$ ms. The $ 40 $ MHz band is composed of different LTE channels with different bandwidths. Hereafter, the findings for a $ 5 $ MHz channel lies between $ 2653.4 $ and $ 2658.4 $ MHz are considered. The ON and OFF states are found using ED with a probability of false alarm of $1\times 10^{-3}$. The received signal is found to have an SNR of $\gamma_0 > 16 $ dB. Therefore, there is no practical risk of miss detecting some of the signals. After performing the detection, the ON and OFF temporal lengths are fitted to exponential distributions mixture with the methodology explained in the following subsection.

\subsection{Exponential distributions mixture fitting}\label{sec:hyper_exp}

The idea of exponential distributions mixture fitting is to fit different parts of the distribution into different independent exponential distributions linearly combined with different weights. Mathematically, for a random variable, $\theta$, the exponentials mixture distribution probability density function (pdf), $ f(\theta) $, is expressed as
  \begin{equation}
  f(\theta)=\sum\limits_{i=1}^{k}w_{i}\lambda_{i} e^{-\lambda_{i}\theta},
  \label{eq:dist_mix}
  \end{equation}  
  where $w_{i}$ is the weight of the $i^{th}$ distribution, $\lambda_{i}$ is its corresponding exponent parameter and $\sum\limits_{i=1}^{k}w_{i} = 1$.
 
Following is the procedure of finding the distribution parameters as the essence of \cite{hyper_exp}. The values of $\lambda_i $ can be chosen to satisfy $ \lambda_{1} < \lambda_{2}< ... < \lambda_{k}, \forall i$. By this, the distribution characterized by the exponent parameter $  \lambda_1$ will decay the slowest compared to the other distribution while the one characterized by $  \lambda_k$ will decay the fastest. Therefore, we can assume that some part of the tail is exclusively characterized by the first exponential distribution. Accordingly,
\begin{subequations} \label{eq:fitting_prop}
\begin{equation}
1 - w_{1}e^{-\lambda_{1}c_1} = F(c_{1}),
\label{eq:fitting_prop1}
\end{equation}
and
\begin{equation}
\sum\limits_{i=2}^{k} w_{i}e^{-\lambda_{i}\theta} = 0 \mbox{ for } \theta > c_1,
\label{eq:fitting_prop2}
\end{equation}
\end{subequations}
where $ c_{1} $ is a sufficiently large value located in the tail of the empirical cumulative distribution function (CDF), $F(\theta) $. In the same way, if $b$ is a constant that satisfies $b>1$, then
\begin{subequations} \label{eq:fitting_prop_b}
\begin{equation}
1 - w_{1}e^{-\lambda_{1}bc_{1}} = F(bc_{1}),
\label{eq:fitting_prop1_b}
\end{equation}
and
\begin{equation}
\sum\limits_{i=2}^{k} w_{i}e^{-\lambda_{i}\theta} = 0 \mbox{ for } \theta > bc_1.
\label{eq:fitting_prop2_b}
\end{equation}
\end{subequations}
Subsequently, $\lambda_1 \mbox{ and }w_1$ are obtained by
\begin{subequations}
\begin{equation}
\lambda_{1} = \dfrac{1}{(b-1)c_{1}}\mbox{ln}\left( \dfrac{1 - F(c_{1})}{1 - F(bc_{1})} \right),
\label{eq:lam1}
\end{equation}
\begin{equation}
w_{1} = \left(1 - F(c_{1})\right)^{\lambda_{1}c_{1}}.
\label{eq:w1}
\end{equation}
\end{subequations}
After finding $\lambda_1 \mbox{ and }w_1$  the pairs $(\lambda_i,w_i)$ for $2 \le i < k$ are found recursively as
\begin{subequations}
\begin{equation}
\lambda_{i} = \dfrac{1}{(b-1)c_{i}}\mbox{ln}\left( \dfrac{1 - F(c_{i})}{1 - F_{i}(bc_{i})} \right),
\label{eq:lami}
\end{equation}
\begin{equation}
w_{i} = 1 - F_{i}(c_{i})e^{\lambda_{i}c_{i}},
\label{eq:wi}
\end{equation}
\end{subequations}
where 
\begin{equation*}
c_i  = c_{1}a^{-(i-1)}, a > b,
\end{equation*}
\begin{equation*}
  F_{i}(c_i) = F_{i-1}(c_i) + \sum\limits_{j=1}^{i-1}e^{-\lambda_{j}c_{i}} ,
 \end{equation*}
 \begin{equation*}
  F_{i}(bc_i) = F_{i-1}(bc_i) + \sum\limits_{j=1}^{i-1}e^{-\lambda_{j}bc_{i}}  ,
  \end{equation*}
   and 
  \begin{equation*}
  F_{1}(\theta) = F(\theta).
 \end{equation*}
Finally, the last pair $(\lambda_k,w_k)$ is found as 
\begin{subequations}
\begin{equation}
w_{k} = 1-\sum\limits_{j=1}^{k-1}w_{j},
\label{eq:wk}
\end{equation}
\begin{equation}
\lambda_{k} = \dfrac{1}{c_{k}}\mbox{ln}\left( \dfrac{w_{k}}{1 - F_{k}(c_{k})} \right).
\label{eq:lamk}
\end{equation}
\end{subequations}
Setting the values of $ c_{1} $, $ b $, and $ a  $  is explained in \cite{hyper_exp}.

As stated in Subsection \ref{sec:Markov}, ON and OFF temporal lengths are assigned the RVs $ x $ and $ y $ respectively. Suppose that both $x$ and $y$ are fitted to $ k $ exponential distributions mixture as
\begin{subequations}
\begin{equation}
  f(x)\approx \sum\limits_{i=1}^{k}w^{x}_{i}\lambda^{x}_{i} e^{-\lambda^{x}_{i}x},
  \label{eq:dist_mix_x}
  \end{equation}
  \begin{equation}
    f(y)\approx \sum\limits_{i=1}^{k}w^{y}_{i}\lambda^{y}_{i} e^{-\lambda^{y}_{i}y},
    \label{eq:dist_mix_y}
    \end{equation}
\end{subequations}
where\\
$ w^{x}_{i} $ is the weight for the exponential distribution number $ i $ for the RV $x$,\\
$ \lambda^{x}_{i} $ is the exponent parameter for the exponential distribution number $ i $ for the RV $x$,\\
$ w^{y}_{i} $ is the weigh for the exponential distribution number $ i $ for the RV $y$,\\
$ \lambda^{y}_{i} $ is the exponent parameter for the exponential distribution number $ i $ for the RV $y$,\\
From the exponential distributions parameters, the channel utilization factor $ u $ is found as
\begin{equation}
u = \dfrac{\sum\limits_{i = 1}^{k} \left(\dfrac{w^{x}_{i}}{\lambda^{x}_{i}}\right)}{\sum\limits_{i = 1}^{k} \left(\dfrac{w^{x}_{i}}{\lambda^{x}_{i}}\right) + \sum\limits_{i = 1}^{k} \left(\dfrac{w^{y}_{i}}{\lambda^{y}_{i}}\right)}
\label{eq:u}
\end{equation}
With a periodic sensing performed each $T$ seconds, the captured opportunities $\zeta$ are found using
\begin{equation}
    \zeta = (1- u) \cdot \left ( \sum\limits_{i=1}^{k}\dfrac{w^{x}_{i}}{\lambda^{x}_{i}T}\left(1-\mathrm{e}^{-\lambda^{x}_{i}T} \right )\right).
    \label{eq:cop}
    \end{equation}  
   \begin{flushright}  $\blacksquare$ \end{flushright}
   \textit{\textbf{Proof}}: see Appendix A.

\subsection{Measurements and fitting results}\label{sec:meas_res}

\begin{figure}[h!]
    \centering
    \includegraphics[width=0.49\textwidth]{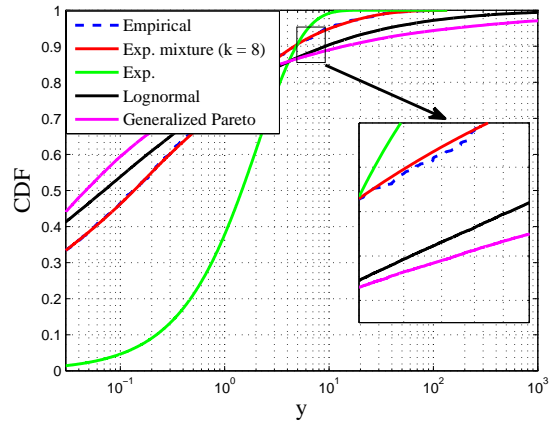} 
    \caption{Empirical and fitted CDF for different distributions. The data is for the OFF periods length in the period of Tuesday, October , 01, 2013, 12:00 - 16:00.}
    \label{fig:diss_dist}
    \end{figure}

To investigate the advantage of using exponential distributions mixture over some other heavy tail distributions, the measurements findings for $ x $ and $ y $ are fitted for different  distributions including exponential, log-normal, generalize Pareto and exponentials mixture. The empirical and fitted distributions for $y$ in the period Tuesday, October , 01, 2013, 12:00 - 16:00, are shown as samples of the results in Fig. \ref{fig:diss_dist}. Choosing the OFF periods and the time is done arbitrarily as a representative case of the results. As Fig. \ref{fig:diss_dist} shows, fitting with a mixture of eight exponential distributions outperforms all other fitted distributions. The goodness of fit is quantitatively evaluated using the log-likelihood estimation defined as \cite{AIC}
\begin{equation}
\Phi(\theta|g(\theta)) = \int f(\theta)\mbox{log}\left(\dfrac{g(\theta)}{f(\theta)}\right)d\theta,
\label{eq:log_like}
\end{equation} 
 where $ \Phi(\theta|f(\theta)) $ is the log-likelihood estimation for a random variable $ \theta $ having an empirical pdf  $f(\theta) $ fitted to a distribution with a pdf $ g(\theta)$. In the case of exponential mixture fitting, $ \Phi(\theta|f(\theta)) $ is found as 
 \begin{equation}
 \Phi(\theta|g(\theta)) = \int f(\theta)\mbox{log}\left(\dfrac{\sum\limits_{i=1}^{k}w_{i}\lambda_{i} e^{-\lambda_{i}\theta}}{f(\theta)}\right)d\theta.
 \label{eq:log_like_exp}
 \end{equation} 
  
Table \ref{tab:log_like} shows the log-likelihood estimation for different distributions. The log-likelihood estimation is calculated for fitting both $ x $ and $ y $ in the same period as the one used to generate Fig.\ref{fig:diss_dist}. 

    \begin{table}[!h]
          \renewcommand{\arraystretch}{1.3}
          \caption{Log-likelihood for fitting different distributions. The fitted date is for the ON periods length, $ x $, and the OFF periods length, $ y $, in the period of Tuesday, October , 01, 2013, 12:00 - 16:00. }
          \label{tab:log_like}
          \centering
          \begin{tabular}{c c | c c}
          \hline\hline
          Distribution    &    & $ \Phi(x|f(x)) $ & $ \Phi(y|f(y)) $ \\
          \hline
          Exponential     &    & $0.368$ & $0.377$  \\
          Lognormal       &   & $ 0.017 $ & $0.013$ \\
          Generalized Pareto & & $ 0.032 $ & $0.029$ \\
          \hline
          \multirow{7}{*} {Exponentials mixture} &  $ k = 2$  & $ \textcolor{blue}{0.292} $ & $\textcolor{red}{0.244}$ \\
                                         &  $ k = 3 $  & $\textcolor{blue}{0.220}$ & $\textcolor{red}{0.228}$\\
                                         &  $ k = 4 $  & $\textcolor{blue}{0.136}$ & $\textcolor{red}{0.131}$\\
                                         &  $ k = 5 $  & $\textcolor{blue}{0.068}$ & $\textcolor{red}{0.066}$\\
                                         &  $ k = 6 $  & $\textcolor{blue}{0.064}$ & $\textcolor{red}{0.063}$\\
                                         &  $ k = 7 $  & $\textcolor{blue}{0.020}$ & $\textcolor{red}{0.019}$\\
                                         &  $ k = 8 $  & $\textbf{\textcolor{blue}{0.005}}$ & $\textbf{\textcolor{red}{0.004}}$\\
          \hline\hline
          \end{tabular}
          \end{table}



\section{Throughput driven Sensing}\label{sec:thrput}

\subsection{MBS-FBS Interference Model}\label{sec:thrput_infer}

\begin{figure*}[h!]
  \centering
  \includegraphics[width=0.7\textwidth]{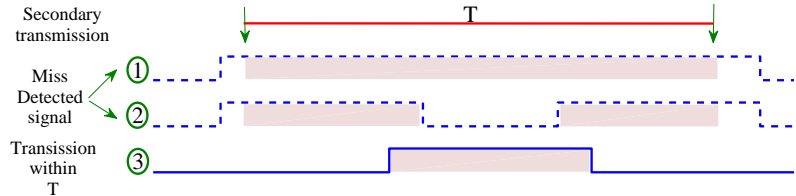} 
  \caption{Mutual operation cases between primary and secondary users. Case 1 and 2 are due to miss detection while case 3 is due to primary transmission reappearance. The binary representation is for the primary user channel occupancy. The solid lines plot represents a detectable signal while the dashed lines plot corresponds to a miss-detected signal.}
  \label{fig:interfer_instances}
  \end{figure*}

Taking into account that the FBS has a very low transmission power compared to the MBS, the interference generated by the mutual operation is assumed to be in one direction, from the MBS to the FBS. Generally, in the spectrum sensing based secondary access framework, depending on the sensing finding, the mutual operation can happen in three cases exhibited in Fig. \ref{fig:interfer_instances} and described below.

\begin{enumerate}

\item \textit{Case 1}: The secondary user miss-detects the primary signal which is active for a time longer than $T$. Accordingly, the secondary user starts to use the channel simultaneously with the primary user.
\item \textit{Case 2}: The secondary user miss-detects the primary signal as in case 1. However, the primary user changes its activity within a period of $T$ one or more times.
\item \textit{Case 3}: An absence of a primary signal is correctly detected and secondary transmission took place accordingly. However, after a time less than $T$, the primary user resumes its operation and possibly changes its activity for once or more.
\end{enumerate}

  Due to the high transmission power of the MBS, the probability of miss-detection is practically zero which eliminates the occurrence of case 1 and 2. Accordingly,  the fraction of time for mutual operation, call it $\tau$, is computed as  
  \begin{equation}
    \tau = \dfrac{1}{u  T}(1-p_{fa})\left(F_{y}(T)\right)u T = (1-p_{fa})\left( 1-\sum\limits_{i=1}^{k}w^{y}_{i}\mathrm{e}^{\left(-\lambda^{y}_{i}T\right)}\right).
    \label{eq:tau_final}
    \end{equation}
   
To calculate the interference due to mutual operation between the MBS and the FBS, the interfering power from the MBS needs to be calculated. In \cite{3GPP_PL}, a model for propagation loss and shadow fading in LTE systems is developed. This model calculates the outdoor-to-indoor loss (O-I), $PL_{M}$, as
  \begin{equation}
  PL_{M} = 36.7 \mbox{log}_{10}(R) + 26 \mbox{log}_{10}(f_{c}) + 0.5 d + 42.7 ,
  \label{eq:PL_M}
  \end{equation}
with $ R $, $ d $ and $ f_c $ denoting the spatial distance between the MBS and the building containing the terminal, the indoor distance between the wall and the terminal and the carrier frequency respectively. The shadow fading is modeled as a zero-mean log-normal distributed RV with a standard deviation $ \sigma_{M} $, Therefore, the received signal from the MBS, $ P^{M}_{r}$, is log-normally distributed as   
  \begin{equation}
  10 \mbox{log}_{10}\left(P_{r}^{M}\right) : \mathcal{N}( \mu_{M}, \sigma_{M}),
  \label{eq:shadow_fading_M}
  \end{equation}
where $ \mathcal{N}( \mu, \sigma) $ denotes a Gaussian distributed RV with a mean of $ \mu $ and a standard deviation of $ \sigma $, $ \mu_{M} = \left(P_{t}^{M} - PL_{M} \right) $ and $ P_{t}^{M} $ is the MBS transmission power.
  
 To calculate the signal-to-interference ratio received at the terminal, the received power from the FBS needs to be obtained. Similarly to the model for the prorogation loss for the MBS signal, in \cite{3GPP_PL}, a model for the indoor path loss for the FBS signal, $ PL_{F} $, is found as  
  \begin{equation}
  PL_{F} = 43.3 \mbox{log}_{10}(d) + 20 \mbox{log}_{10}(f_{c})+ 11.5,
  \label{eq:PL_F}
  \end{equation}
   with a zero mean log-normal shadow fading having a standard deviation $ \sigma_{F} $. Consequently, the FBS received power at the terminal $ P_{r}^{F} $ is distributed as   
  \begin{equation}
  10 \mbox{log}_{10}\left(P_{r}^{F}\right) : \mathcal{N}( \mu_{F}, \sigma_{F}),
  \label{eq:shadow_fading_F}
  \end{equation}
  where $ \mu_{F} = \left(P_{t}^{F} - PL_{F} \right)  $ if $ P_{t}^{M} $ is the FBS transmission power.

  \subsection{Periodic sensing optimization}
  
When the FBS is transmitting in the interference free periods, the achieved FBS throughput, call it $C_0$ is limited by the SNR, $\gamma_0$, as  
  \begin{equation}
  C_0 = B\cdot \mbox{log}_{2}(1+\gamma_0).
  \label{eq:C0}
  \end{equation}
  In the same way, the FBS throughput in the interfered periods, $ C $, is
  \begin{equation}
    C = B\cdot \mbox{log}_{2}(1+\gamma),
    \label{eq:C}
    \end{equation}  
where $ \gamma $ is the signal to interference plus noise ratio (SINR). For the FBS throughput during the whole operation time, call it $C_{all}$, $C_0$ and $C$ are weighted by $(1-\tau)$ and $\tau$ respectively and summed up. Moreover, it should be noted that the FBS is efficiently transmitting with a factor of $\eta$ during a fractional time of captured opportunities, $\zeta_{s}$, (see (\ref{eq:eff}) and (\ref{eq:zeta_sys})). Hence $C_{all}$ is expressible as   
   \begin{equation}
   C_{all} = \left(\eta \cdot \zeta_{s}\right) \left(\tau \cdot C + (1-\tau) \cdot C_0\right).
   \label{eq:C_all}
   \end{equation}
   
Let us define $\chi$ as the ratio between the drop in the average FBS throughput due to the opportunistic access and the interference free average FBS throughput. Accordingly, $\chi$ is found as
  \begin{equation}
  \chi = 1-\dfrac{E\{C_{all}\}}{E\{C_{0}\}}.
  \label{eq:chi}
  \end{equation}
The mutual operation fractional time, $ \tau $, and the throughput drop, $ \chi $, are related as
  \begin{equation}
  \tau = \left(1-\dfrac{1-\chi}{\eta \cdot \zeta_{s}} \right)  \left(\dfrac{\mu_{F}-\sigma_{n}^{2}}{\mu_{M}-\sigma_{n}^{2}}\right).
  \label{eq:tau_formula}
  \end{equation}
  \begin{flushright}  $\blacksquare$ \end{flushright}  
  \textit{\textbf{Proof}}: See Appendix B.\\
As (\ref{eq:tau_final}) and (\ref{eq:tau_formula}) are equivalent, equalizing their right hand sides yields
\begin{equation}
\chi = 1-\eta \cdot \zeta_{s} \left( 1-\alpha \cdot \tau \right)
\label{eq:chi_T_form}
\end{equation}
where
\begin{equation*}
\alpha = \dfrac{\mu_{M}-\sigma_{n}^{2}}{\mu_{F}-\sigma_{n}^{2}}
\end{equation*}
Considering that $\eta$ and $\zeta_{s}$ are $T$ dependant as in (\ref{eq:eff}), (\ref{eq:cop}) and (\ref{eq:zeta_sys}). Then, (\ref{eq:chi_T_form}) can be solved for $T$ by specifying the targeted value of the throughput drop, $\chi$. However, specifying the targeted throughput drop, $\chi$, provides \textit{a suboptimal} solution for the periodic sensing interval, $T$, as there may be lower achievable throughput drop with another value of $T$. Moreover, there is no guarantee that there is a valid solution for all values of $\chi$ as there is a minimum throughput drop that can be achieved depending on the MBS traffic. The \textit{optimal} solution for the periodic sensing interval, $T_{opt}$, is achieved by minimizing the throughput drop, $\chi$, as
\begin{equation}
T_{opt} = \underset{T}{\mbox{arg min}}\left\{ 1-\eta \cdot \zeta_{s} \left(1- \alpha \cdot \tau \right) \right\}.
\label{eq:chi_min}
\end{equation}
Hereafter, the highest achievable throughput corresponding to $T_{opt}$ will be denoted as $C_{opt}$.

  \section{Numerical results}\label{sec:res}
  
As stated in Subsection \ref{sec:meas}, the occupancy in a span of $40$ MHz in the $2.6$ GHz LTE band is measured. Two $5$ MHz channels were found in that band when the measurements were conducted. The two $5$ MHz channels show basically the same occupancy behaviour with highly correlated handled traffic during the measurements time. The empirical data and the modeled channel occupancy using exponential distributions mixture of the two channels are used as an input for a simulation study. The simulation part treats the throughput based opportunistic sharing framework developed in this paper. Below Table \ref{tab:sim_par} shows the simulation and model parameters \footnote{The value of $P_{t}^F$ is adjusted to give an average interference free throughput of 100 Mbps.}$^{, }$\footnote{The inter FBS interference is considered as a part of the background noise.}.
  
  \begin{table}[!h]
            \renewcommand{\arraystretch}{1.3}
            \caption{Simulation and model parameters }
            \label{tab:sim_par}
            \centering
            \begin{tabular}{c c }
            \hline\hline
            Parameter        & Value/Description  \\
            \hline
            $p_{fa}$  &  $1 \times 10^{-3}$     \\
            $t_{s}$   &    $ 20 $ ms  \\
            $P_{t}^M$ &  $ 40.0 $ dBm  \\
            $P_{t}^F$  &  $ 23.85 $ dBm \\
            $\sigma_{M} $ &  $ 7 $ dB \cite{3GPP_PL} \\
            $\sigma_{F} $ &  $ 4 $ dB \cite{3GPP_PL}  \\
            $\sigma_{n}^{2} $ &  $ -170 $ dBm/Hz   \\
            $L $ &  $ 2 $ channels\\
            $k $ &  $ 8 $ \\

            \hline\hline
            \end{tabular}
            \end{table}

\begin{figure*}[h!]
    \centering
    \subfloat[]{\label{fig:u_Csl}\includegraphics[width=0.49\textwidth]{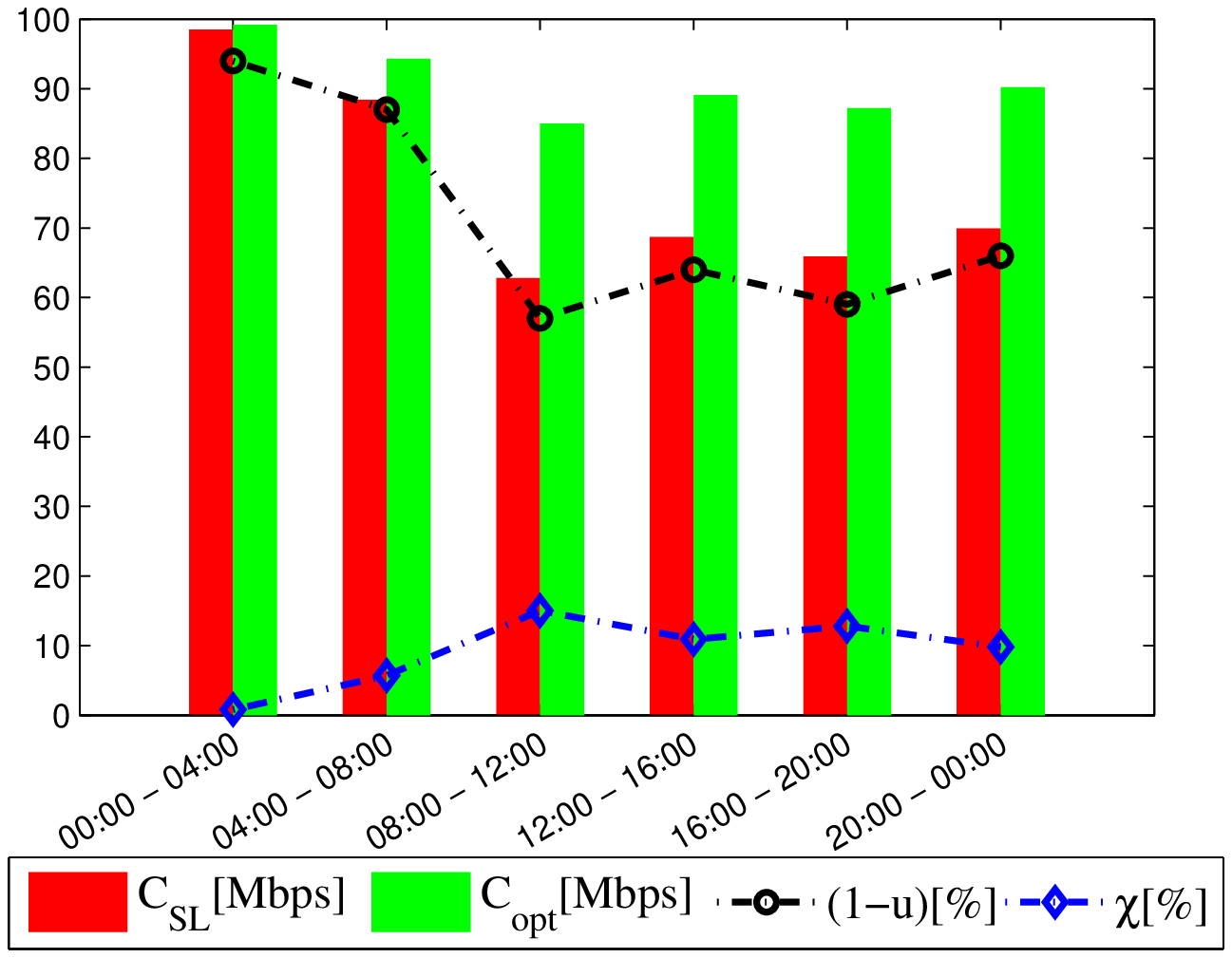}}
    \subfloat[]{\label{fig:Optimal_T}\includegraphics[width=0.49\textwidth]{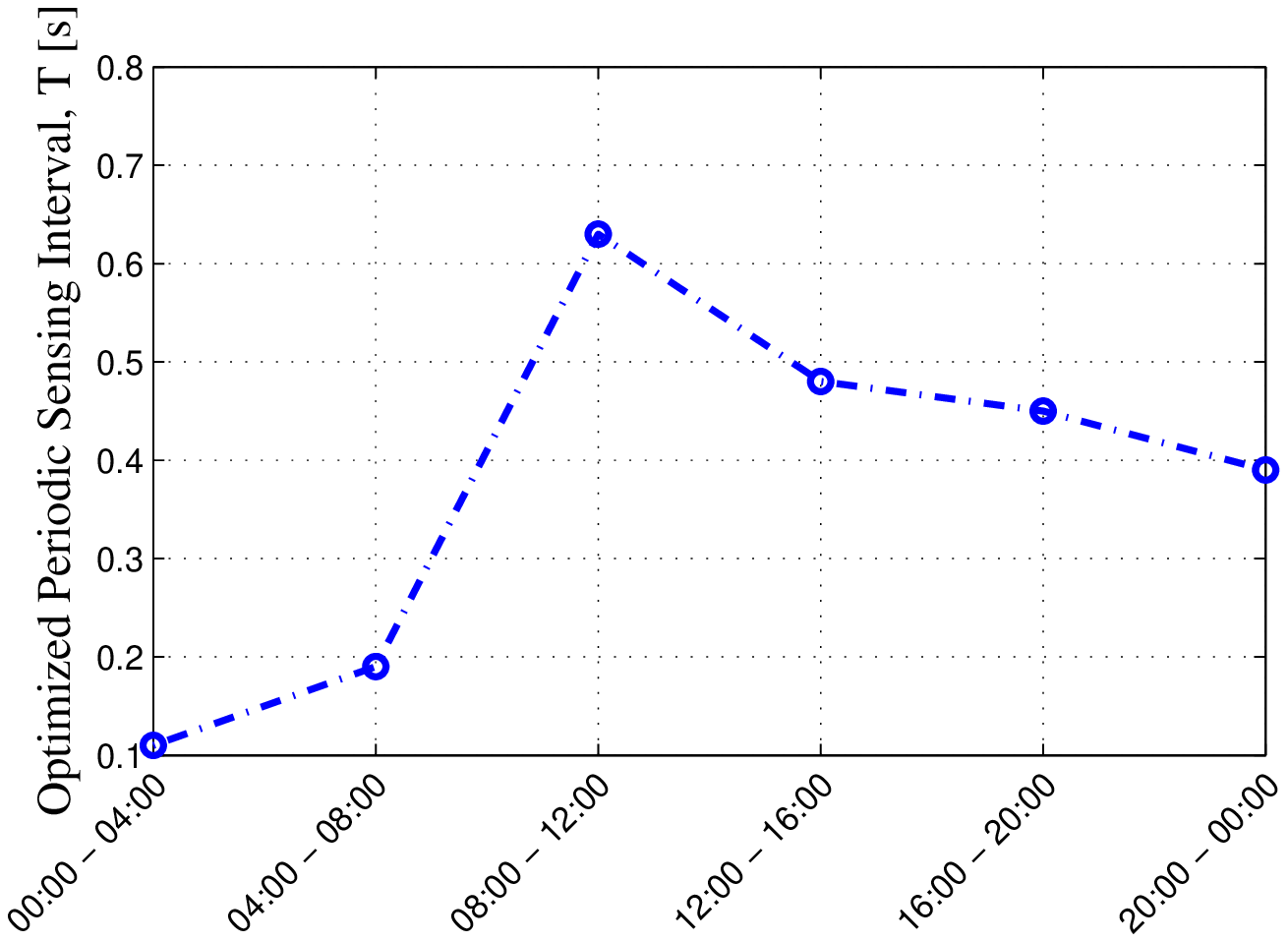}}
    \caption{(a) The senseless throughput, the optimal achieved throughput, the available opportunities for FBS and the minimum achieved interference-free throughput drop. (b) Optimized periodic sensing intervals.}
    \label{fig:U_Csl_T}
    \end{figure*}

For benchmarking purposes, senseless throughput, $C_{SL}$, is defined as the FBS throughput when no sensing takes place. Instead, the FBS randomly chooses one of the available channels for its transmission. The senseless approach is adopted in \cite{Beam_subset}. As a consequence for the senseless strategy, whenever there is an MBS transmission, there exists mutual operation and interfered FBS transmission. Fig. \ref{fig:u_Csl} shows the senseless throughput and the optimal achievable throughput when FBS periodic sensing interval is optimized according to (\ref{eq:chi_min}). The same figure shows the available interference free opportunities for FBS, $(1-u)$. Moreover, the optimal throughput, $C_{opt}$, is shown in Fig. \ref{fig:u_Csl} with the corresponding optimized periodic sensing intervals in Fig. \ref{fig:Optimal_T}. All the results shown in Fig. \ref{fig:U_Csl_T} are obtained using the measurements data taken in different periods of the day October, 02, 2013. This day has been randomly picked. As depicted by Fig. \ref{fig:u_Csl}, the more the interference free opportunities, the higher the senseless throughput, the lower the optimal interference free throughput drop and the lower the optimal periodic sensing interval. In accordance to that, the gain in the achieved throughput when periodic sensing interval is optimized reaches its peak with the lowest available opportunities. This result reflects the necessity of applying the periodic sensing intervals optimization as increasing the throughput is needed more with the increase off the traffic which is expected to happen as the same time for both MBS and FBSs.

\begin{figure}[h!]
      \centering
      \includegraphics[width=0.49\textwidth]{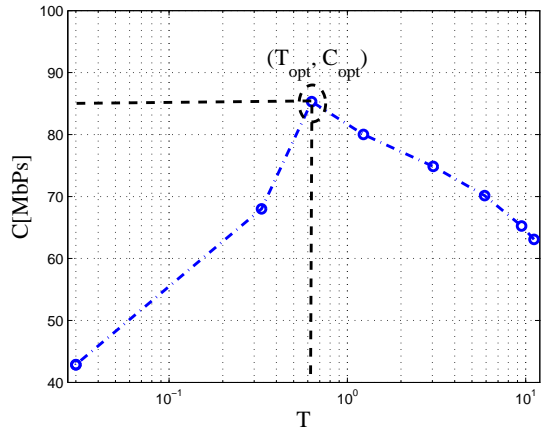} 
      \caption{The change of the achieved throughput with the change of the periodic sensing interval, $T$. The results are for the period October, 02, 2013, 08:00 - 12:00.}
      \label{fig:T_thrput}
      \end{figure}

Fig. \ref{fig:T_thrput} shows an example of the achievable throughput with different values of the periodic sensing interval. The shown results are for the measurements performed during the period October, 02, 2013, 08:00 - 12:00. As the figure shows, the highest obtainable throughput, $C_{opt}$, of about $ 85 $ Mbps is attained when the optimal periodic sensing interval is used. For the other values of the periodic sensing interval, lower or higher than the optimal periodic sensing interval, the throughput is lower than $C_{opt}$. The figure illustrates the credibility of using the optimal periodic sensing interval.

\begin{figure}[h!]
    \centering
    \includegraphics[width=0.49\textwidth]{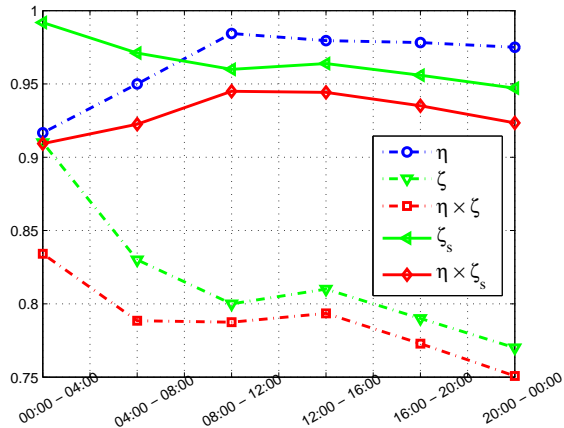} 
    \caption{Obtained values of the transmission efficiency, $\eta$, the captured opportunities, $\zeta$ and the utilized opportunities for a single channel and the whole system.}
    \label{fig:eta_zeta}
    \end{figure}

Fig. \ref{fig:eta_zeta} shows the transmission efficiency, $\eta$, and the captured opportunities, $\zeta$, for a single channel. The figure illustrates the contradictory trends for the transmission efficiency and the captured opportunities, i.e. longer periodic sensing interval results in higher transmission efficiency but more missed opportunities. The total efficiently used opportunities for transmission are then decided by the product of the two quantities which is shown in the figure. For the whole system consideration, the captured opportunities in all channels, $ \zeta_{s} $ is then taken into account instead of the captured opportunities in each channel separately. Consequently, the total efficiently used opportunities determined by the product of $\eta$ and $ \zeta_{s} $ is higher for our two channels system as the figure exhibits.
  
\begin{figure*}[h!]
  \centering
    \subfloat[]{\label{fig:thrput_drop}\includegraphics[width=0.49\textwidth]{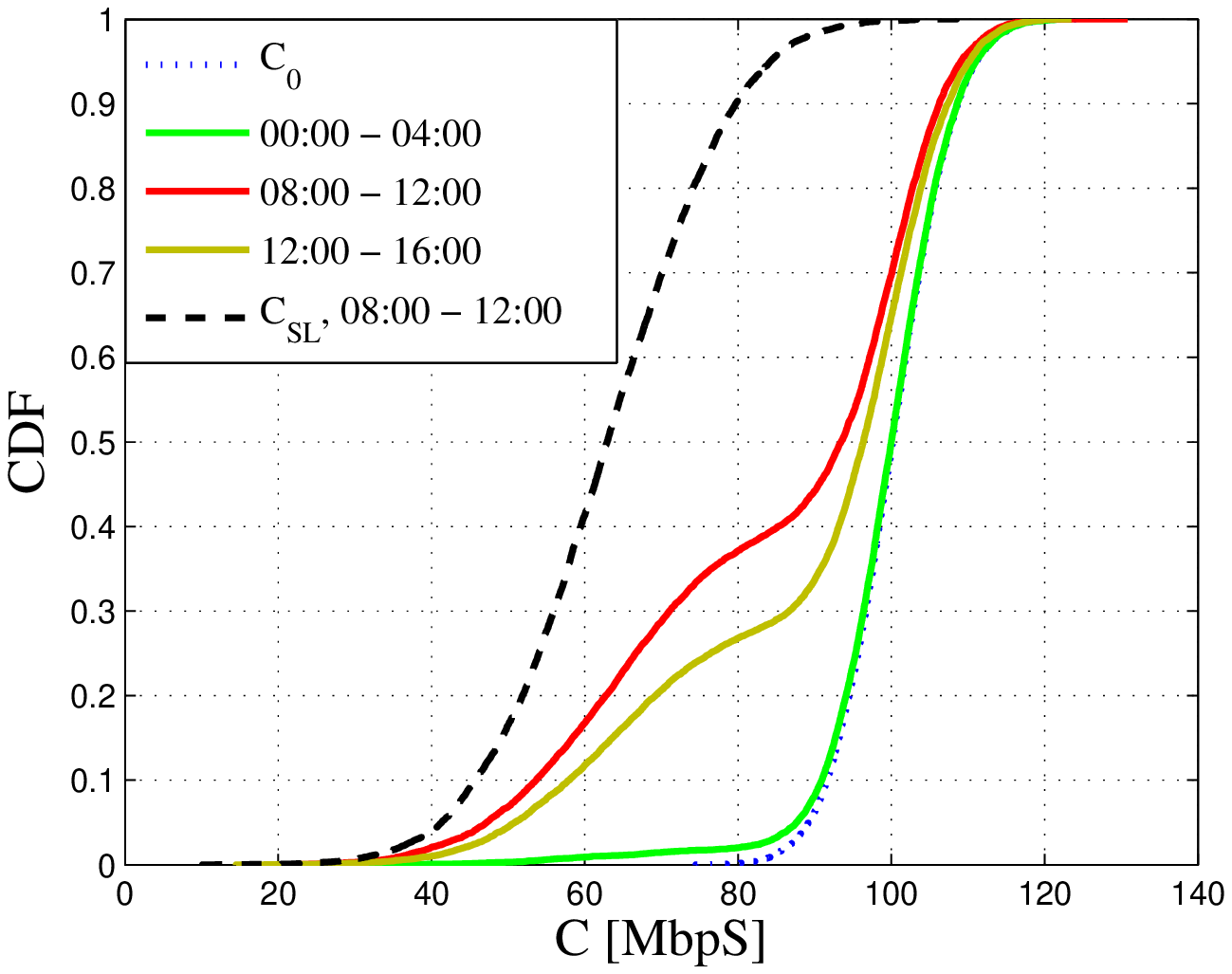}}
    \subfloat[]{\label{fig:thrput_drop_8_12}\includegraphics[width=0.49\textwidth]{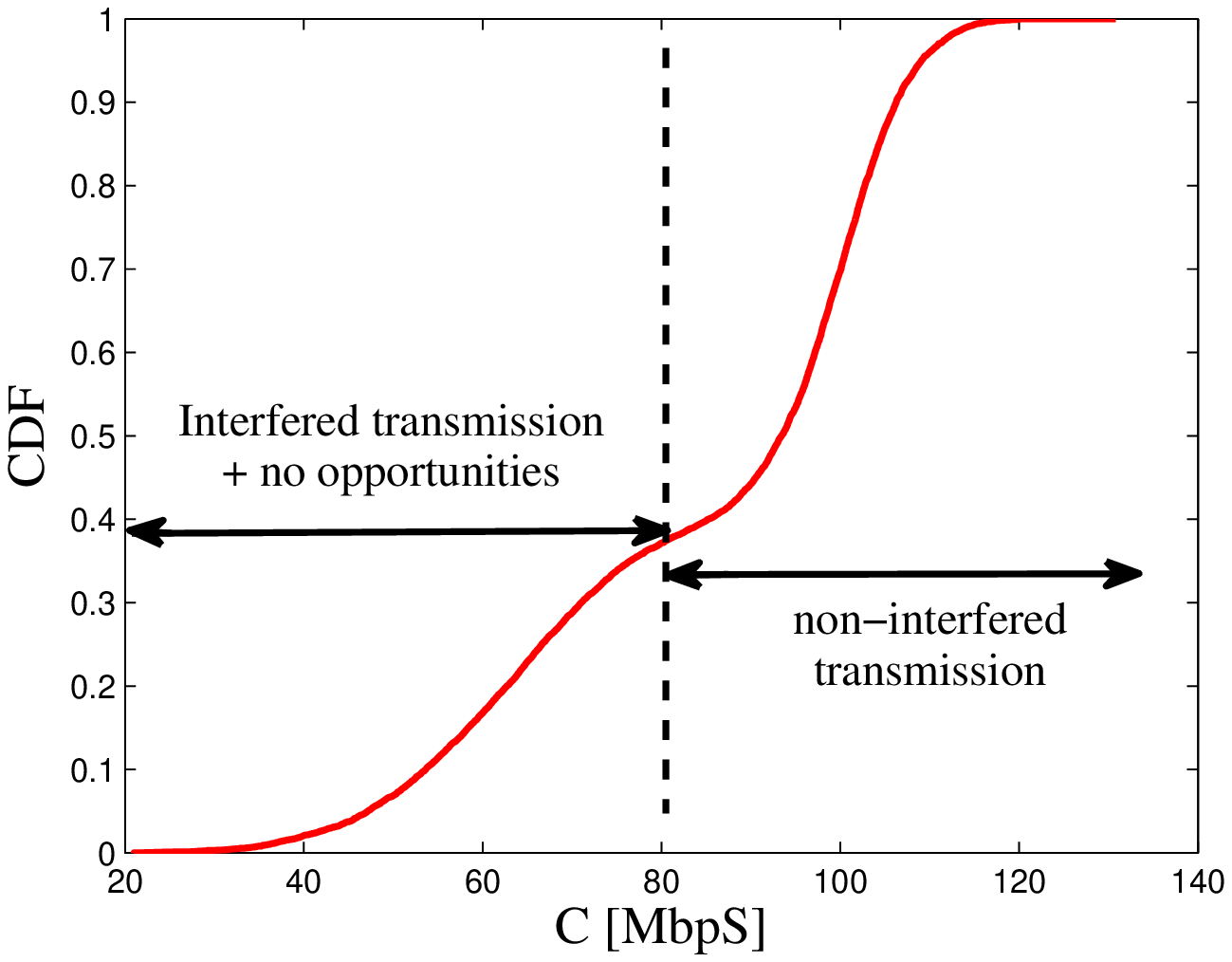}}

    \caption{(a) The CDF for the throughput in the cases of MBS traffic absence, sense less scenario and periodic sensing optimization. (b) Example of the regions of the throughput CDF }

    \end{figure*}

Fig. \ref{fig:thrput_drop} shows the CDF of the throughput when the periodic sensing is optimized. The CDF curves are bounded between two limits. The upper limit is the case of MBS traffic absence and noise only existence, i.e. $C_{0}$. The lower limit is the senseless throughput, $C_{SL}$. The lower limit is traffic dependant. Therefore, the period of the highest traffic is taken as the absolute lower limit for the whole day of measurements. Apart from $C_{0}$ and $C_{SL}$, the throughput CDF curves in Fig. \ref{fig:thrput_drop} can be divided into two regions. The first region is the interfering transmission plus no opportunities region. In this region the throughput is degraded due to one of two reasons. The first reason is a simultaneous FBS and MBS transmission caused by MBS transmission reappearance while FBS is being transmitting. The second reason is that the sensing has resulted in no opportunities to utilize by the FBS. The second region is the non-interfered transmission region where the opportunities are being used by the FBS without MBS transmission reappearance. Fig. \ref{fig:thrput_drop_8_12} demonstrates the two regions of the throughput CDF for the period October, 02, 2013, 08:00 - 12:00 as an example.

\begin{figure}[h!]
        \centering
        \includegraphics[width=0.49\textwidth]{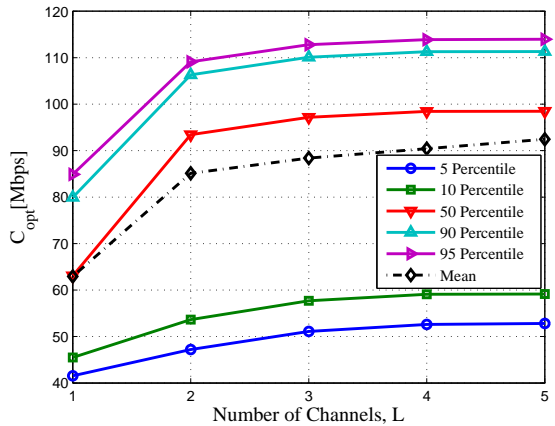} 
        \caption{$5, 10, 50, 90 \mbox{ and }95$ percentile of the throughput when different number of channels are available.}
        \label{fig:cap_perc}
        \end{figure}

The influence of the number of the available channels, $L$, on the achieved optimal throughput is studied and the results are presented in Fig \ref{fig:cap_perc} for the period October, 02, 2013, 08:00 - 12:00. As there existed two $5$ MHz channels in the measured band, the data for higher number of channels is a synthetic data generated by bootstrapping the data for these two channels and other channels with different bandwidths. The mean and the $5, 10, 50, 90 \mbox{ and }95$ percentiles of the throughput when the number of the available channels changes from $1$ to $5$ are shown in Fig \ref{fig:cap_perc}. The results when a single channel is available basically represent the senseless scenario which is the worst case scenario. The diversity gain in the throughput is then judged by the increase of the throughput with the increase of the available number of channels. In this regard, it is seen from Fig. \ref{fig:cap_perc} that increasing the number of channels would increase the achieved optimal throughput. However, for more than three channels the optimal throughput saturates. This trend is explained as follows, with the increase of the number of channels, the captured opportunities found as in (\ref{eq:zeta_sys}) would increase. Yet, this increase in the captured opportunities impact grows less significantly with increasing the number of available channels till it saturates at $L = 4$ channels.

  \section{Conclusions}\label{sec:conc}
  Cognitive LTE femto-cells whose access the LTE macro-cell spectrum on an opportunistic basis are considered in this paper. The paper contributes in introducing a framework for downlink channels access for LTE femto-cells. The channel access framework is based on employing periodic sensing using energy detection for finding the free channels. The framework objective is to maximize the LTE femto-cells downlink throughput. The throughput is maximized by optimizing the sensing intervals in a multi-channel spectrum sharing system by compromising the transmission efficiency, the captured opportunities and the interference from the macro-cell base station. The statistics of the macro-cell channel occupancy impacts the optimization procedure. This channel occupancy is empirically modeled as a part of the paper contribution. The analytical derivations have shown that there exist an optimal periodic sensing interval which maximizes the femto-cell downlink throughput. Hence, a more efficient use of the spectrum assigned for LTE downlink transmission is achievable following the framework presented in this paper.
  
  \renewcommand{\theequation}{A.\arabic{equation}}
    \setcounter{equation}{0}  
    
    \renewcommand{\thefigure}{A.\arabic{figure}}
        \setcounter{figure}{0}  
    
   \section*{\begin{center}
       Appendix A\\
       Proof of equation(\ref{eq:cop})
       \end{center}}
   
\begin{figure}[h!]
         \centering
         \includegraphics[width=0.49\textwidth]{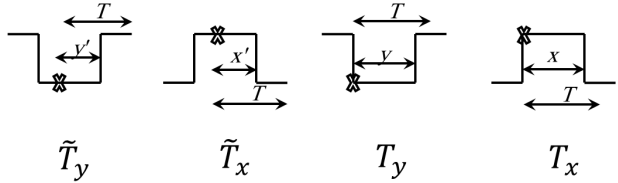} 
         \caption{Possible channel statuses at sensing}
         \label{fig:AppendixA}
         \end{figure}

Let $T_{z}$ to be the average opportunities during a period of $T$. $z$ is either $x$ or $y$ denoting if the channel is occupied or free. The sensing point can be an end or a start of an idle period, so in that case we use $\tilde{T}_{z}(t)$ instead of $T_{z}$. Therefore, we have four possible cases, as illustrated in Fig. \ref{fig:AppendixA}.
    
Let $y^{'}$ to denote the remaining time of an OFF period when sensing is performed. The distribution of $y^{'}$ is given by \cite{cox}
\begin{equation}
F_{Y}(y^{'}) = \dfrac{1-F_{Y}(y^{'})}{E\{y\}}.
\label{eq:App_1_1}
\end{equation}
In the same way, if the remaining time of an ON period when sensing is performed is denoted as $x^{'}$, then $x^{'}$ is distributed as
\begin{equation}
F_{X}(x^{'}) = \dfrac{1-F_{X}(x^{'})}{E\{X\}}.
\label{eq:App_1_2}
\end{equation}
Renewal theory is applied to find $ T_{y}(t), T_{x}(t), \tilde{T}_{y}(t) \mbox{ and } \tilde{T}_{x} (t) $. Following are the derived formulas.
\begin{equation}\label{eq:app3}
T_{y}(t) = t \int\limits_{t}^{\infty}\dfrac{1-F_{Y}(y)}{E\{y\}}dy + \int\limits_{0}^{t}\dfrac{1-F_{Y}(y)}{E\{y\}}\left(y + \tilde{T}_{x}(t-y)\right)dy.
\end{equation}
\begin{equation}\label{eq:app4}
T_{x}(t) = \int\limits_{0}^{t}\dfrac{1-F_{X}(x)}{E\{x\}}\tilde{T}_{y}(t-x)dx.
\end{equation}
\begin{equation}\label{eq:app5}
\tilde{T}_{y}(t) = t \int\limits_{t}^{\infty}f_{Y}(y)dx + \int\limits_{0}^{t}f_{Y}(y) \left( y + \tilde{T}_{x}(t-y)\right)dy.
\end{equation}
\begin{equation}\label{eq:app6}
\tilde{T}_{x}(t) = \int\limits_{0}^{t}f_{X}(x) \tilde{T}_{y}(t-x)dx.
\end{equation}
If Laplace transform is performed for (\ref{eq:app3}), (\ref{eq:app4}), (\ref{eq:app5}), and (\ref{eq:app6}), then following formulas are obtained
\begin{equation}
E\{y\} \cdot T_{y}(s) = \dfrac{F_{Y}(s)-F_{Y}(0)}{s^{2}} + \left(1-F_{Y}(s)\right)\tilde{T}_{x}(s).
\end{equation}
\begin{equation}
E\{x\} \cdot T_{x}(s) =  \left(1-F_{x}(s)\right)\tilde{T}_{y}(s).
\end{equation}
\begin{equation}
\tilde{T}_{x}(s) =  f_{X}(s)\tilde{T}_{y}(s).
\end{equation}
\begin{equation}
\tilde{T}_{y}(s) =  \dfrac{f_{Y}(0)-f_{Y}(s)}{s^{2}} + f_{Y(s)}\tilde{T}_{x}(s).
\end{equation}
Therefore,
\begin{equation}
T_{y}(s) = \dfrac{1}{E\{y\}\cdot s^{2}}\cdot \left( F_{Y}(s)-F_{Y}(0)  \right).
\end{equation}
\begin{equation}
T_{x}(s) = \dfrac{1-F_{X}(s)}{E\{x\}\cdot s^{2}} \cdot \left(\dfrac{f_{Y}(0)-f_{Y}(s)}{1-f_{Y}(s)f_{X}(s)} \right).
\end{equation}
If we define the missed opportunities, $\Upsilon$, as the fraction of missed OFF periods, then, $\Upsilon$ is expressed as $ \Upsilon= \Upsilon_{y} + \Upsilon_{x}, $. $\Upsilon_{y}$ representing the opportunities missed from the start of an OFF period till the sensing is performed which is formulated as
\begin{equation}
\Upsilon_{y} = (1-u)\cdot \left( \dfrac{1}{T} \int\limits_{0}^{T}\dfrac{1-F_{Y}(y)}{E\{y\}}\tilde{T}_{x}(T-y)dy\right).
\end{equation}
On the other hand, $\Upsilon_{x}$ is the opportunities missed due to changing the status of the channel from ON to OFF within a period of $T$. $\Upsilon_{x}$ is found as
 \begin{equation}
 \Upsilon_{x} = u \cdot \left( \dfrac{T_{x}(T)}{T}\right).
 \end{equation}
Sine the total sum of the missed opportunities, $\Upsilon$ and the captured opportunities, $\zeta$, is equivalent to the total available opportunities, then
 \begin{equation}
  \zeta = (1-u) - \Upsilon.
  \end{equation}
Therefore, with the exponential mixture of the ON and OFF periods shown in (\ref{eq:dist_mix_x}) and (\ref{eq:dist_mix_y}), $\zeta$ is found using
\begin{equation}
  \zeta = (1- u) \cdot \left ( \sum\limits_{i=1}^{k}\dfrac{w^{x}_{i}}{\lambda^{x}_{i}T}\left(1-\mathrm{e}^{-\lambda^{x}_{i}T} \right )\right).
  \end{equation}  

\renewcommand{\theequation}{B.\arabic{equation}}
    \setcounter{equation}{0}  
   \section*{\begin{center}
     Appendix B\\
     Proof of equation(\ref{eq:tau_formula})
     \end{center}}

At the terminal served by the FBS, having the received power from the FBS, the interfering power from the MBS and the noise variance results in a received SINR, $\gamma$, calculated as
     \begin{equation}
    \gamma = \dfrac{P_{r}^{F}}{P_{r}^{M} + \sigma_{n}^{2} }.
    \label{eq:SINR}
     \end{equation}
     Cellular systems are interference dominating systems where the noise variance is neglectable compared to the interference \cite{Ben_boundLR}. Therefore, with the log-normal distribution of both $P_{r}^{F}$ and $P_{r}^{M}$ as in (\ref{eq:shadow_fading_M}) and (\ref{eq:shadow_fading_F}), $\gamma$ is approximated as a log-normal distributed RV as     
     \begin{equation}
      10 \mbox{log}_{10}\left(\gamma\right) : \mathcal{N}( \mu_{\gamma}, \sigma_{\gamma}^{2}).
     \end{equation}     
where $\mu_{\gamma}  = \mu_{F} - \mu_{M}$ and $\sigma_{\gamma}^{2}  = \sigma_{F}^{2} + \sigma_{M}^{2}$.\\
In the same way, $ \gamma_{0} $ is expressed as 
 \begin{equation}
     \gamma_{0} = \dfrac{P_{r}^{F}}{\sigma_{n}^{2}},
     \label{eq:SNR}
      \end{equation}
Accordingly, $\gamma_{0}$ is a log-normally distributed RV as 
\begin{equation}
      10 \mbox{log}_{10}\left(\gamma_{0}\right) : \mathcal{N}( \mu_{F}-\sigma_{n}^{2}, \sigma_{F}^{2}).
      \label{eq:SNR_dist}
     \end{equation}
Noting that all the quantities in (\ref{eq:SNR_dist}) are taken in their logarithmic (dB) scale.\\
From (\ref{eq:C0}), (\ref{eq:C}), (\ref{eq:C_all}) and (\ref{eq:chi}),
\begin{equation}
       \begin{split}
      1-\chi &= \eta \cdot \zeta_{s} \dfrac {E\Big\{\tau B \mbox{log}_{2}\left(1+\gamma \right)  + (1-\tau) B \mbox{log}_{2}\left(1+\gamma_{0} \right) \Big\}}{E\Big\{B \mbox{log}_{2}\left(1+\gamma_{0}\right)\Big\}}\\
            &= \eta \cdot \zeta_{s} \dfrac {B \cdot E\Big\{\tau \cdot 10 \mbox{log}_{10}\left(1+\gamma \right)  + (1-\tau) \cdot 10 \mbox{log}_{10}\left(1+\gamma_{0} \right) \Big\}}{B \cdot E\Big\{10 \mbox{log}_{10}\left(1+\gamma_{0}\right)\Big\}}
     \end{split}
           \label{eq:proof2}
\end{equation}     
In the serving area of the FBS, the valid assumption of $ \gamma,\gamma_{0} \gg 1 $ leads to
  \begin{equation}
        \begin{split}
         1-\chi &= \eta \cdot \zeta_{s} \dfrac {E\Big\{\tau \cdot 10 \mbox{log}_{10}\left(\gamma \right)  + (1-\tau) \cdot 10 \mbox{log}_{10}\left(\gamma_{0} \right) \Big\}}{E\Big\{10 \mbox{log}_{10}\left(\gamma_{0}\right)\Big\}}\\
                &= \eta \cdot \zeta_{s} \dfrac{\tau \cdot (\mu_{F} - \mu_{M} ) + (1-\tau)\cdot (\mu_{F} - \sigma_{n}^{2})}{(\mu_{F} - \sigma_{n}^{2})} 
       \end{split}
             \label{eq:proof3}
           \end{equation}
           Therefore,
           \begin{equation}
             \tau = \left(1-\dfrac{1-\chi}{\eta \cdot \zeta_{s}} \right)  \left(\dfrac{\mu_{F}-\sigma_{n}^{2}}{\mu_{M}-\sigma_{n}^{2}}\right).
             \end{equation}

  \bibliographystyle{IEEEtran}
  \bibliography{biblo}


  \end{document}